\documentclass[11pt,a4paper]{article}

\pdfoutput=1

\usepackage{jheppub}
\usepackage[normalem]{ulem}
\usepackage{slashed}
\usepackage{tikz-feynman}
\usepackage{bm}
\usepackage{hyperref}

\usepackage{caption,subcaption}

\def\beq{\begin{equation}}
\def\eeq{\end{equation}}
\def\bea{\begin{eqnarray}}
\def\eea{\end{eqnarray}}
\def\bma{\begin{matrix}}
\def\ema{\end{matrix}}

\def\nl{\nonumber\\}

\def \cB{{\cal B}}

\def \cO{{\cal O}}
\def \cA{{\cal A}}

\def \cH{{\cal H}}

\def \tha{\theta}

\def \la{\lambda}

\def \cH{{\cal H}}

\def \({\left(}
\def \){\right)}
\def \[{\left[}
\def \]{\right]}

\def \ok{\overline{K}^0}
\def \arg{\rm arg}

\newcommand{\ket}[1]{\left|#1\right\rangle}

\newcommand{\Braket}[1]{\left\langle #1\right\rangle}

\allowdisplaybreaks

\begin{document}

\title{\boldmath Flavor $SU(3)$ in Cabibbo-favored $D$-meson decays}

\author[a]{Bhubanjyoti Bhattacharya,}
\author[b]{Alakabha Datta,}
\author[c]{Alexey A Petrov,}
\author[b]{and John Waite}

\affiliation[a]{Department of Natural Sciences, Lawrence Technological University, Southfield, MI 48075, USA}
\affiliation[b]{Department of Physics and Astronomy, \\
108 Lewis Hall, University of Mississippi, Oxford, MS 38677-1848, USA}
\affiliation[c]{Department of Physics and Astronomy\\
        Wayne State University, Detroit, MI 48201, USA}
        \emailAdd{bbhattach@ltu.edu}
\emailAdd{datta@phy.olemiss.edu}
\emailAdd{apetrov@wayne.edu}
\emailAdd{jvwaite@go.olemiss.edu}

\abstract{
Model-independent description of nonleptonic decays of charmed mesons is a challenging task due to
large nonperturbative effects of strong interactions on the transition amplitudes. We discuss the
equivalence of two different flavor-$SU(3)$-based descriptions of Cabibbo-favored non-leptonic decays
of charmed mesons to two-pseudoscalars final states including the $\eta$ and $\eta^\prime$ mesons.
}


\arxivnumber{2107.13564}

\preprint{
\begin{flushright}
WSU-HEP-2102
\end{flushright}
}

\maketitle

\section{Introduction}

Studies of non-leptonic decays of charmed mesons constitute a primary method of investigations into direct CP-violation
in that system. Even though the experimental precision for studying $D$ decays has steadily improved over the past decade,
theory calculations have faced severe challenges. Precise numerical predictions of CP-violating observables are not possible
at the moment due to large non-perturbative contributions from strong interactions affecting weak-decay amplitudes. A way out
in such a situation involves phenomenological fits of decay amplitudes to experimentally measured decay widths of charmed
mesons. If the number of fit parameters is smaller than the number of experimentally measured observables, then predictions
are possible. Such fits require a defined procedure on how to parametrize complex-valued decay amplitudes \cite{Petrov:2021idw}.

One way to approach the problem is to note that the light-quark operators in the weak effective Hamiltonian governing heavy-quark decays,
as well as the initial and final states form product representations of a flavor $SU(3)_F$ group. These product representations can be
reduced with the help of the Wigner-Eckart theorem. This way, a basis is chosen, in which all decay amplitudes can be expanded in terms
of the reduced matrix elements. Such an approach was applied to both $B$-decays \cite{Savage:1989ub,Grinstein:1996us,He:2018joe,He:2018php} and
$D$-decays \cite{Quigg:1979ic,Kaeding:1995vq,Savage:1991wu,He:2018joe,He:2018php}. In the limit of exact $SU(3)_F$ symmetry, all decays of a triplet of
$D$-mesons, $D^0, D^+,$ and $D^+_s$, into two light octet meson states can be parametrized in terms of five independent complex
parameters \cite{Quigg:1979ic}. We will refer to this approach as the ``$SU(3)_F$ matrix-elements approach.''

Alternatively, a topological flavor-flow approach can be used. Developed in the study of $B$-decays \cite{Gronau:1994rj,Buras:1998ra,Imbeault:2006ss,He:2018joe,He:2018php}, it has been applied to the charm sector \cite{Bhattacharya:2012pc,Bhattacharya:2012ah,He:2018joe,He:2018php}. The flavor-flow approach postulates a basis of universal flavor
topologies for various decay amplitudes.\footnote{We remind the reader that while the flavor-flow diagrams do resemble Feynman graphs, they are not computed in perturbative field theory due to large non-perturbative QCD effects.} $SU(3)_F$ symmetry can be used to relate
decay amplitudes, as both the light-quark final states and initial $D$-mesons transform under it. These universal topological amplitudes can be
fitted to the existing experimental data. Due to long-distance effects, particularly rescattering among hadronic final states, often multiple flavor-flow
diagrams contribute to the same process. A subset of {\it linear combinations} of flavor-flow amplitudes can then be identified as the basis set for the
flavor-flow approach.

The two approaches described above are equivalent if the number of reduced matrix elements in the $SU(3)_F$ matrix-elements approach is
 equal to the number of diagrammatic combinations in the flavor-flow approach, both describing the same set of decay amplitudes. Such
 an equivalence has been shown in the of exact $SU(3)_F$ symmetry \cite{Gronau:1994rj,Zeppenfeld:1980ex}, as well as when first-order
 $SU(3)_F$-breaking corrections are taken into account \cite{Muller:2015lua}. Here we revisit the question of equivalence of the two descriptions
 and discuss the fit quality of the available data in both approaches.

Non-leptonic decays of charmed mesons can be additionally classified according to the rate of suppression of the (leading-order) weak-decay amplitudes by the Wolfenstein parameter, $\la = \sin\theta \simeq 0.2$ \cite{Wolfenstein:1983yz}, where
$\theta$ is the Cabibbo angle. Such amplitudes may contain zero, one, or two powers of $\la$. Weak-hadronic decays of charm are, therefore, categorized into Cabibbo-favored (CF) decays ($\cA \propto V^*_{cs}V^{}_{ud} \sim \cO(1)$), singly-Cabibbo-suppressed (SCS) decays ($\cA \propto V^*_{cq}V^{}_{uq} \sim \cO(\la)$ where $q = d, s$), and doubly-Cabibbo-suppressed (DCS) decays
($\cA\propto V^*_{cd}V^{}_{us} \sim \cO(\la^2))$. Since such classification is external to QCD, both the flavor $SU(3)_F$ and topological flavor-flow approaches can, in principle, be used to parametrize all CF, SCS, and DCS amplitudes. This can be considered as an advantage, as some fit parameters can be obtained from the CF and/or DCS transitions and then used to predict CP-violating asymmetries in SCS decays
\cite{Golden:1989qx,Pirtskhalava:2011va,Cheng:2012wr,Cheng:2012xb,Feldmann:2012js}.
This is so because the quark-level transitions for CF ($c\to su{\bar d}$) and DCS ($c\to du{\bar s}$) modes involve four distinct quark flavors that belong to the first two generations and therefore do not generate CP-asymmetries in the Standard Model at leading order in $\lambda$. To execute this program one needs to control $SU(3)_F$-breaking corrections in both
approaches \cite{Kaeding:1995vq,Savage:1991wu,Muller:2015lua,Hiller:2012xm,Falk:2001hx}.

Here we take a different look at the equivalency of the flavor $SU(3)_F$ and the topological flavor-flow approaches. Since the Wolfenstein parameter $\lambda$ is external to any QCD-based parametrization of decay amplitudes, one can, theoretically, dial any value for it. In particular, setting $\lambda=0$ would only leave CF decays as experimental data for a fit. It is interesting to note that in this case, in the $SU(3)_F$ limit, we would be left with {\it three} irreducible $SU(3)_F$ amplitudes and {\it four} topological flavor-flow amplitudes. In this paper we explore the equivalency of the phenomenological descriptions of CF charmed-meson decays in light of this discrepancy.

This paper is organized as follows. In Section \ref{sec:fits} we review both the flavor $SU(3)_F$ and topological flavor-flow
approaches to CF charm decays. We extend the discussion by including CF decays with the real $\eta$ and $\eta^\prime$ states
and present associated fits. In Section \ref{sec:connections}, we discuss the connections between those two approaches.
We conclude in Section \ref{sec:conclusions}.

\section{Cabibbo-favored decays in light of flavor-$SU(3)$ symmetry }
\label{sec:fits}

$SU(3)_F$ symmetry plays a prominent role in both the $SU(3)_F$ matrix-elements and topological flavor-flow approaches. Both methods use the fact that the
initial and final states transform under some product representations of the $SU(3)_F$ group. In particular, the initial state $D$-mesons,
$\ket{D^0}, \ket{D^+}, \ket{D^+_s}$, form a triplet of $SU(3)_F$, while the nine pseudoscalar mesons ($\pi^\pm,\pi^0,K^\pm,K^0,\overline{K}^0,\eta,\eta'$)
contain both an octet and a singlet. The two approaches differ by the choice of the ``basis'' parameters, which we discuss below.

In what follows, we employ physical $\eta$ and $\eta'$ states that are constructed from the $SU(3)$ octet $\eta_8$ and the singlet $\eta_1$ states using octet-singlet mixing,
\bea
\eta &=& -\cos\tha\,\eta_8 - \sin\tha\,\eta_1 ,
\nonumber \\
\eta' &=& -\sin\tha\,\eta_8 + \cos\tha\,\eta_1,
\eea
where the octet $\eta_8$ and the singlet $\eta_1$ states are defined as
\bea
\eta_8 &=& (u{\bar u} + d{\bar d} - 2\,s{\bar s})/\sqrt{6},
\nonumber \\
\eta_1 &=& (u{\bar u} + d{\bar d} + s{\bar s})/\sqrt{3} ,~
\eea
and $\tha$ is the $\eta-\eta'$ mixing angle. This mixing angle has nothing to do with weak decays of heavy flavors and can be fixed externally, for example from $B$-meson decays \cite{Datta:2001ir} or radiative decays of $J/\psi$ \cite{Petrov:1997yf} into $\eta$ and $\eta^\prime$ final states. Thus, we do not consider it a fit parameter. Oftentimes, we will use $\tha = \arcsin(1/3)$.

\subsection{$SU(3)_F$ matrix-elements approach}

The $SU(3)_F$ matrix-elements approach uses the fact that the Hamiltonian governing $D$ decays into the light mesons also transforms as a product representations of $SU(3)_F$. The quark-level Hamiltonian for CF $D$-meson decays can be written as
\beq
{\cH}_\text{CF}~=~\frac{G_F}{\sqrt{2}}V_{ud}V_{cs}^*(\bar ud)(\bar sc) + {\rm h.c.}
\label{eq:HCF}
\eeq
We begin by considering the Wigner-Eckart decompositions of the CF $D\to PP$ amplitudes using $SU(3)_{F}$ symmetry. An element of $SU(3)$ can be
represented using the state $\ket{\bm{r}YII_3}$ where $\bm{r}$ is the irreducible
representation (irrep) of the state, $Y$ is its hypercharge, while $I$ and $I_3$ stand for the isospin and its
third component, respectively. Under $SU(3)_{F}$ symmetry, the light quarks $u$, $d$, and $s$ (and the respective antiquarks)
transform as the fundamental triplet (anti-triplet) represented by,
\bea
\ket{u}~=~\ket{\bm{3},\frac{1}{3},\frac{1}{2},\frac{1}{2}}\,,\quad
\ket{d}~=~\ket{\bm{3},\frac{1}{3},\frac{1}{2},-\frac{1}{2}}\,,\quad
\ket{s}~=~\ket{\mathbf{3},-\frac{2}{3},0,0}\,, \\
\ket{\overline{u}}~=~-\ket{\bm{\overline{3}},-\frac{1}{3},\frac{1}{2},-\frac{1}{2}}\,,\quad
\ket{\overline{d}}~=~ \ket{\bm{\overline{3}},\frac{1}{3},\frac{1}{2},\frac{1}{2}}\,,\quad
\ket{\overline{s}}~=~\ket{\bm{\overline{3}},\frac{2}{3},0,0}\,.
\eea
The charm quark (and anti-quark) is heavy and transforms as an $SU(3)_{F}$ singlet represented by $\ket{\bm{1},0,0,0}$.
Using this notation one can show that the CF Hamiltonian in Eq.~(\ref{eq:HCF}) contains the irreps
$\bm{\overline{15}}$ and $\bm{6}$ \cite{Quigg:1979ic,Savage:1991wu,Kaeding:1995vq} and can be represented by,
\beq
\mathcal{H}_\text{CF}~=~ -~\frac{G_F}{\sqrt{2}}V_{ud}V_{cs}^*\(A\,{\cO}^{(\bm{\overline{15}})}_{\frac{2}{3},1,-1} + C\,{\cO}^{(\bm{6})}_{\frac{2}{3},1,-1}\) + {\rm h.c.}\,,
\eeq
where we have used the notation $O^{(\bm{r})}_{Y,I,I_3}$ to denote the $SU(3)$ operators, whereas $A$ and $C$ represent
their respective coefficients.

As mentioned previously, the final states transform under a product representation of $SU(3)_F$. Since the octet-octet final states must respect
Bose symmetry, we only consider the following products of $SU(3)_F$ irreps,

\bea
\[(\bm{8} + \bm{1}) \times (\bm{8} + \bm{1})\]_{PP} &=& (\bm{8}\times\bm{8})_{\rm sym}
+ (\bm{8}\times\bm{1}) + \bm{1}\,,~
\nonumber \\
&=& {\bm 27} + {\bm 8}_{{\bm 8}\times{\bm 8}} + {\bm 8}_{{\bm 8}\times{\bm 1}} + \bm{1}_{{\bm 8}\times{\bm 8}}
+ \bm{1}\,.
\label{eq:FinStDec}
\eea
Note that there are two octets in Eq.~(\ref{eq:FinStDec}): one from the octet-octet final state and the other from the
octet-singlet one.

Now, of the above irreps only the $\bm{27}$ and $\bm{8}$ appear in the products $\bm{\overline{15}}\times\bm{3}$
and $\bm{6}\times\bm{3}$ needed to construct the states $\ket{\cH|D}$. Furthermore, $\bm{\overline{15}}\times\bm{3}$
contains both a $\bm{27}$ and an $\bm{8}$, while $\bm{6}\times\bm{3}$ contains only an $\bm{8}$. Therefore, it appears
that $D\to PP$ amplitudes can be represented using the following three independent reduced matrix elements.
\beq
A_{27} ~=~ \Braket{\bm{27}|\cO^{\bm{\overline{15}}}|\bm{3}}\,, \quad
A_8    ~=~ \Braket{\bm{8}|\cO^{\bm{\overline{15}}}|\bm{3}}\,, \quad
C_8    ~=~ \Braket{\bm{8}|\cO^{\bm{6}}|\bm{3}}\,.
\label{eq:SU3me}
\eeq
These reduced matrix elements depend on five real parameters -- three magnitudes and two relative strong phases (one overall phase can be ignored).
The amplitudes for the CF $D\to PP$ processes can be constructed using these reduced matrix elements. As there are two different octets in
Eq.~(\ref{eq:FinStDec}), in general this would imply two additional reduced matrix elements for the $\cO^{\bm{\overline{15}}}$ and $\cO^{\bm{6}}$ operators,
$A_8^{(1)}$ and $C_8^{(1)}$ respectively. In Section \ref{sec:connections} we will show that indeed in order to get a complete description of these decays
one must include these additional matrix elements that correspond to the $SU(3)_F$-singlet final state.

Assuming them to be the same, $A_8^{(1)}=A_8$ and $C_8^{(1)}=C_8$, which can be motivated by a nonet symmetry, the final states containing physical
$\eta$ and $\eta^\prime$ contain an admixture of singlet and octet $SU(3)_F$ amplitudes. The decay amplitudes into those final states can be written as
shown in Table ~\ref{tab:CFeta}.
\begin{table}[!h]
\begin{center}
\begin{tabular}{|c|c|}\hline
Decay&Representation\\
\hline
$D^0\to \ok\eta$&$\frac{1}{10\sqrt{3}}\left[(3A_{27}-2A_8+\sqrt{10}C_8)\cos\theta+2(\sqrt{10}A_8-5C_8)\sin\theta\right]$\\
$D^0\to\ok\eta'$&$\frac{1}{10\sqrt{3}}\left[(3A_{27}-2A_8+\sqrt{10}C_8)\sin\theta-2(\sqrt{10}A_8-5C_8)\cos\theta\right]$\\
\hline
$D_s^+\to\pi^+\eta$&$\frac{1}{5\sqrt{3}}\left[(3A_{27}-2A_8-\sqrt{10}C_8)\cos\theta-(\sqrt{10}A_8+5C_8)\sin\theta\right]$\\
$D_s^+\to\pi^+\eta^\prime$&$\frac{1}{5\sqrt{3}}\left[(3A_{27}-2A_8-\sqrt{10}C_8)\sin\theta+(\sqrt{10}A_8+5C_8)\cos\theta\right]$\\
\hline
\end{tabular}
\caption{$\eta-\eta^\prime$ decay-amplitude representations with $A_8^{(1)}=A_8$ and $C_8^{(1)}=C_8$
in the $SU(3)_F$ matrix-elements approach.}
\label{tab:CFeta}
\end{center}
\end{table}
Assuming, for simplicity, $\tha = \arcsin(1/3)$, all CF decay amplitudes can be written in terms of only three complex parameters of Eq.~(\ref{eq:SU3me}).
We provide a representation of the decay amplitudes in terms of those parameters in Table~\ref{tab:CF}.
\begin{table}[!h]
\begin{center}
\begin{tabular}{|c|c|}\hline
Decay & $SU(3)_F$ Amplitude \\\hline
$D^0\to K^-\pi^+$ & $\frac{G_F}{\sqrt{2}}V_{ud}V_{cs}^*~\frac{1}{5}\left(\sqrt{2}A_{27}+\sqrt{2}A_8-\sqrt{5}C_8\right)$ \\
$D^0\to \ok\pi^0$ & $\frac{G_F}{\sqrt{2}}V_{ud}V_{cs}^*~\frac{1}{10}\left(3A_{27}-2A_8+\sqrt{10}C_8\right)$ \\
$D^0\to\ok\eta$  & $\frac{G_F}{\sqrt{2}} V_{ud}V_{cs}^* \, \frac{1}{15\sqrt{3}} \left(3\sqrt{2}A_{27} + \sqrt{2}(\sqrt{5}-2) A_8-\sqrt{5}(\sqrt{5}-2)C_8\right)$ \\
$D^0\to\ok\eta'$  & $\frac{G_F}{\sqrt{2}}V_{ud}V_{cs}^* ~\frac{1}{30\sqrt{3}}\left(3A_{27}-2(1+4\sqrt{5})A_8+\sqrt{10}(1+4\sqrt{5}) C_8\right)$ \\ \hline
$D^+\to \ok\pi^+$ &$\frac{G_F}{\sqrt{2}}V_{ud}V_{cs}^*~\frac{1}{\sqrt{2}}A_{27}$ \\ \hline
$D^+_s\to\ok K^+$ & $\frac{G_F}{\sqrt{2}}V_{ud}V_{cs}^*~\frac{1}{5}\left(\sqrt{2}A_{27}+\sqrt{2}A_8+\sqrt{5}C_8\right)$ \\
$D^+_s\to\pi^+\eta$ & $\frac{G_F}{\sqrt{2}}V_{ud}V_{cs}^* ~\frac{1}{15\sqrt{3}}\left(6\sqrt{2}A_{27} - \sqrt{2} (4+\sqrt{5}) A_8 - \sqrt{5}(4+\sqrt{5})C_8\right)$ \\
$D^+_s\to\pi^+\eta'$&$\frac{G_F}{\sqrt{2}}V_{ud}V_{cs}^*~\frac{1}{15\sqrt{3}}\left(3A_{27}+2(2\sqrt{5}-1)A_8
+\sqrt{10}(2\sqrt{5}-1) C_8\right)$ \\ \hline
\end{tabular}
\end{center}
\caption{$SU(3)_F$ matrix-elements representation of Cabibbo-Favored Decays in the Standard Model. Note that
the $\eta-\eta^\prime$ angle $\tha = \arcsin(1/3)$.}
\label{tab:CF}
\end{table}
These matrix elements can be fit to experimentally-measured branching ratios.

The measured branching ratios, $\cB$, for the CF $D\to PP$ decays are given in Table~\ref{tab:CFBR}. The absolute value of each
decay amplitude can be determined from the measured branching ratios using,
\beq
|\cA_{D\to PP}| ~=~ \sqrt{\frac{8\pi\hbar\,m^2_D\,\cB_{D\to PP}}{\tau_D\,p^*}} ,~
\eeq
where $p^*$ refers to the magnitude of the three-momentum of each final-state pseudoscalar in the $D$-meson rest frame.
\begin{table}[h!]
\begin{center}
\begin{tabular}{|c|c|c|}
\hline
 Meson	& Decay		  & Branching Ratio (\%) \\ \hline
$D^0$	&$K^-\pi^+$   &$3.950\pm0.031$ \\
		&$\ok\pi^0$	  &$2.480\pm0.044$ \\
		&$\ok\eta$	  &$1.018\pm0.012$ \\
		&$\ok\eta'$	  &$1.898\pm0.064$ \\ \hline
$D^+$	&$\ok\pi^+$	  &$3.124\pm0.062$ \\ \hline
$D^+_s$	&$\ok K^+$	  &$2.95\pm0.14$ \\
		&$\pi^+\eta$  &$1.70\pm0.09$ \\
		&$\pi^+\eta'$ &$3.94\pm0.25$ \\ \hline
\end{tabular}
\end{center}
\caption{Experimental branching ratios for CF $D$ decays taken from~\cite{Zyla:2020zbs}.
\label{tab:CFBR}}
\end{table}
Since there are eight measured $D\to PP$ branching ratios that depend on five real parameters (three magnitudes and two
relative phases of three reduced matrix elements), a $\chi^2$-minimization fit can be employed to determine the parameters.
Such a fit has three degrees of freedom. We perform a fit by constraining the $C_8$ amplitude to be purely real and find,

\bea
\chi^2_{\rm min}/{\rm dof} &=& 7477/3 ,~ \nl
A_{27} &=& \(0.279 \pm 0.002 \)\,{\rm GeV}^3 ,~ \nl
A_8    &=& \(0.840  \pm 0.008  \)\,e^{(59 \pm 1)^\circ i}\,{\rm GeV}^3 ,~ \nl
C_8    &=& \(0.17 \pm 0.02 \)\,e^{(-58 \pm 2)^\circ i}\,{\rm GeV}^3 .~
\eea
Clearly, the fit is very poor. This leads us to believe that the above description of CF $D\to PP$ decays in terms of
the minimum number of $SU(3)_F$ reduced matrix elements is incomplete and needs to be modified.

In the next section we discuss another parametrization of the same matrix elements, in terms of the topological
flavor-flow amplitudes. We again identify the minimal set of basis amplitudes to describe the CF decays in the flavor-$SU(3)$ limit.
This minimal set appears to work better, seemingly providing an adequate description of CF decays, including those
with the $\eta$ and $\eta^\prime$ mesons in the final state.

\subsection{Topological flavor-flow approach}

\begin{figure}[h]
\centering
  \begin{subfigure}[t]{0.3\textwidth}
      \centering
      \includegraphics[width=\textwidth]{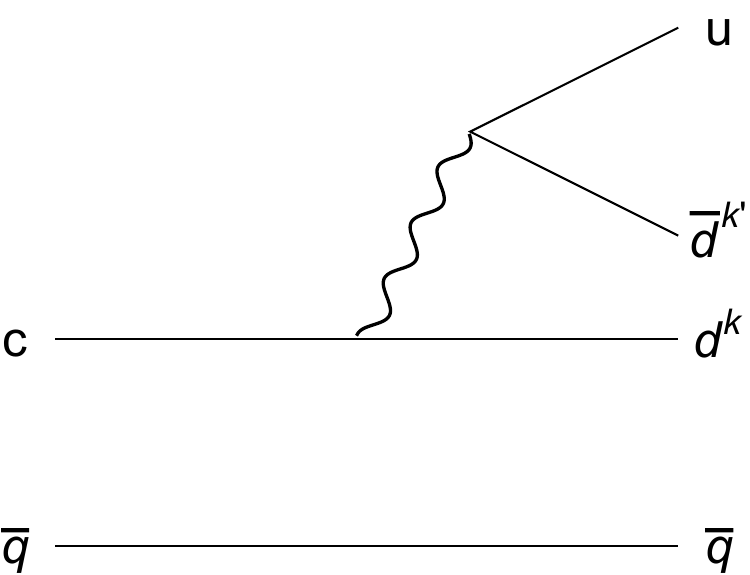}
      \caption{Color-favored tree ($T$)}
  \end{subfigure} \hspace{0.3in}
  \begin{subfigure}[t]{0.3\textwidth}
      \centering
      \includegraphics[width=\textwidth]{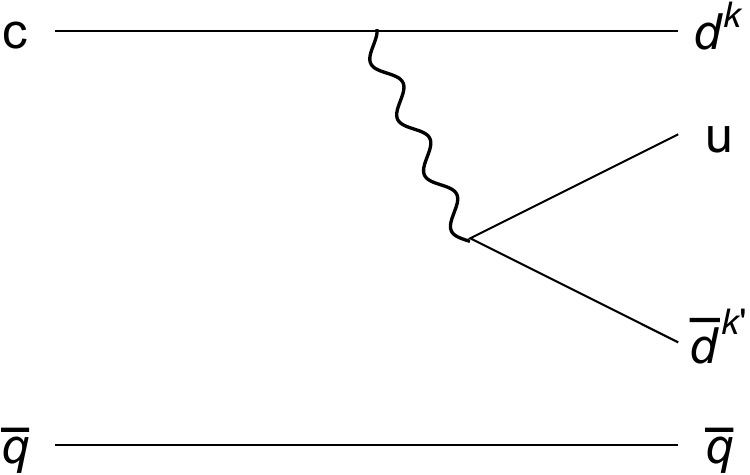}
      \caption{Color-suppressed tree ($C$)}
  \end{subfigure} \vskip 0.3in
  \begin{subfigure}[b]{0.3\textwidth}
      \centering
      \includegraphics[width=\textwidth]{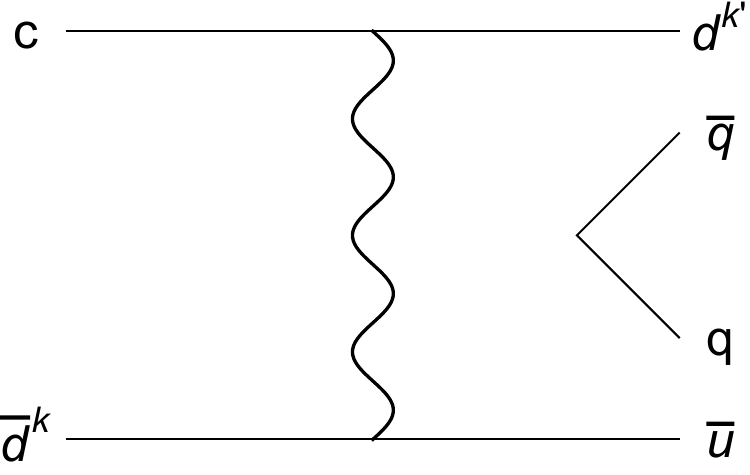}
      \caption{Exchange ($E$)}
  \end{subfigure} \hspace{0.3in}
  \begin{subfigure}[b]{0.3\textwidth}
      \centering
      \includegraphics[width=\textwidth]{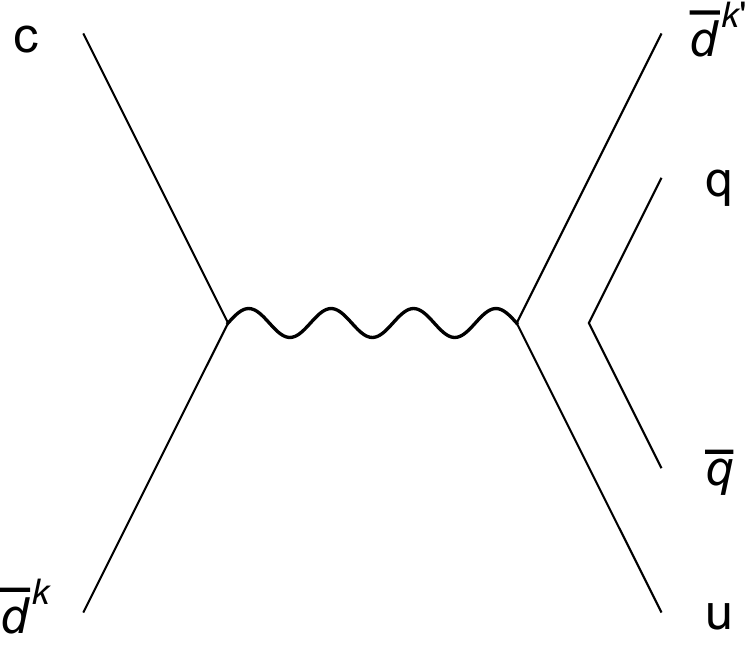}
      \caption{Annihilation ($A$)}
  \end{subfigure}
\caption{Topological flavor-flow diagrams used to describe CF $D\to PP$ decays.}
\label{fig:SMdiagrams}
\end{figure}

The eight CF $D\to PP$ decays can also be described in terms of topological flavor-flow diagrams using
$SU(3)_F$ symmetry, as discussed in Ref.~\cite{Bhattacharya_2010}. Based on the Hamiltonian in Eq.~(\ref{eq:HCF}), the amplitudes
for the CF $D\to PP$ decays can be represented using four flavor topologies shown in Fig.~\ref{fig:SMdiagrams}. The basis of the topological
amplitudes is obtained by identifying the color-favored tree ($T$), color-suppressed tree ($C$), exchange ($E$), and annihilation ($A$) amplitudes.
These four topological amplitudes depend on seven real parameters, four magnitudes and three relative phases (once again one overall
phase is ignored).

The amplitudes for CF decays with at least one $\eta$ or $\eta'$ meson in the final state have explicit dependence
on the $\eta-\eta'$ mixing angle. The topological flavor-flow representation for these decays are given in Table~\ref{tab:arbanglediag}.
\begin{table}[h]
\centering
\begin{tabular}{|c|c|}\hline
 Decay & Representation\\\hline
$D^0\to \ok\eta$	 &$\frac{G_F}{\sqrt{2}}V_{ud}V_{cs}^*~\(\frac{\cos\theta}{\sqrt{6}}(C-E)+\frac{\sin\theta}{\sqrt{3}}(C+2E)\)$\\
$D^0\to\ok\eta'$	 &$\frac{G_F}{\sqrt{2}}V_{ud}V_{cs}^*~\(\frac{\sin\theta}{\sqrt{6}}(C-E)-\frac{\cos\theta}{\sqrt{3}}(C+2E)\)$\\
\hline
$D_s^+\to\pi^+\eta$	 &$\frac{G_F}{\sqrt{2}}V_{ud}V_{cs}^*~\left(\frac{2\cos\theta}{\sqrt{6}}(T-A)-\frac{\sin\theta}{\sqrt{3}}(T+2A)\right)$\\
$D_s^+\to\pi^+\eta$	 &$\frac{G_F}{\sqrt{2}}V_{ud}V_{cs}^*~\left(\frac{2\sin\theta}{\sqrt{6}}(T-A)+\frac{\cos\theta}{\sqrt{3}}(T+2A)\right)$\\
\hline
\end{tabular}
\caption{Amplitudes for $\eta-\eta'$ CF decays in the topological-diagram representation with the assumption that the octet and singlet diagrams are equal.}
\label{tab:arbanglediag}
\end{table}

Once again employing $\tha = \arcsin(1/3)$, in Table~\ref{tab:diagrep}, we express all CF $D\to PP$ decays in terms
of flavor-topological diagrams.
\begin{table}[!h]
\centering
\begin{tabular}{|c|c|}\hline
        Decay & Diagrammatic Amplitude \\\hline
$D^0\to K^-\pi^+$    &$\frac{G_F}{\sqrt{2}}V_{ud}V_{cs}^*~(T+E)$  \\
$D^0\to\ok\pi^0$     &$\frac{G_F}{\sqrt{2}}V_{ud}V_{cs}^*~\frac{1}{\sqrt{2}}(C-E)$   \\
$D^0\to\ok\eta$      &$\frac{G_F}{\sqrt{2}}V_{ud}V_{cs}^*~\frac{1}{\sqrt{3}}C$   \\
$D^0\to\ok\eta'$     &$\frac{G_F}{\sqrt{2}}V_{ud}V_{cs}^*~\(-\frac{1}{\sqrt{6}}\)(C+3E)$  \\ \hline
$D^+\to\ok\pi^+$     &$\frac{G_F}{\sqrt{2}}V_{ud}V_{cs}^*~(C+T)$ \\ \hline
$D^+_s\to\ok K^+$    &$\frac{G_F}{\sqrt{2}}V_{ud}V_{cs}^*~(C+A)$ \\
$D^+_s\to\pi^+\eta$  &$\frac{G_F}{\sqrt{2}}V_{ud}V_{cs}^*~\frac{1}{\sqrt{3}}(T-2A)$ \\
$D^+_s\to\pi^+\eta'$ &$\frac{G_F}{\sqrt{2}}V_{ud}V_{cs}^*~\frac{2}{\sqrt{6}}(T+A)$ \\ \hline
\end{tabular}
\caption{Amplitudes for CF $D$ decays expressed in terms of $SU(3)_F$ flavor-topological diagrams
\label{tab:diagrep}}
\end{table}

The above diagrammatic description of the CF $D\to PP$ processes leads to a parametrization of the eight decay modes in terms of seven real parameters.
This time, with one remaining degree of freedom, a $\chi^2$-minimization fit can once again be performed. We perform such a fit by constraining $T$ to be purely real and find,

\bea
\chi^2_{\rm min}/{\rm dof} &=& 1.36/1 ~, \nl
T &=& (0.366 \pm 0.003)\,{\rm GeV}^3 ~,\nl
C &=& (0.298 \pm 0.002)\,e^{i (-151.0 \pm 0.4)^\circ}~{\rm GeV}^3~, \nl
E &=& (0.201 \pm 0.004)\,e^{i ( 119.3 \pm 0.8)^\circ}~{\rm GeV}^3~, \nl
A &=& (0.04  \pm 0.01 )\,e^{i  (63 \pm 9)^\circ}~{\rm GeV}^3~.
\eea
This fit appears to be excellent, suggesting that the diagrammatic representation of CF $D\to PP$ decays aligns well with
experimental measurements. Note that the diagrammatic approach has one additional complex-valued amplitude (i.e.\ two additional real-valued parameters) compared to the $SU(3)_F$ matrix-elements approach. We observe a significant decrease in the minimum value of $\chi^2$ even though the diagrammatic description is still overdetermined, i.e. there are more observables than parameters.

In the following section we investigate the differences between the two approaches and present an argument for greater consistency between the two parametrizations.

\section{Connections between flavor-flow and matrix elements in $SU(3)_F$}
\label{sec:connections}

An obvious difference between the two approaches presented in the previous section is that, even in the limit of exact $SU(3)_F$ symmetry, the minimal bases contain
different numbers of complex independent parameters: three in the matrix-elements approach and four in the flavor-flow approach. Yet, we expect an equivalence between
the two approaches \cite{Gronau:1994rj,Zeppenfeld:1980ex}. The implication is either that the $SU(3)_F$ matrix-elements approach described above is incomplete or
that the diagrammatic approach has too many parameters. A key observation that follows is that the one-to-one correspondence between group theory and diagrams is
possible to achieve by treating the decays involving only octets separately from those also involving singlets. In order to demonstrate these separate correspondences,
in Table \ref{tab:CFeta8eta1}, we have listed the $SU(3)_F$ matrix-elements and flavor-flow representations side-by-side.
\begin{table}[h!]
\begin{center}
\begin{tabular}{|c|c|c|}\hline
Decay & Matrix Elements & Diagrams \\ \hline
\multicolumn{3}{|c|}{$SU(3)_F$ octet-octet final states} \\ \hline
$D^0\to K^-\pi^+$    &$\frac{1}{5}\(\sqrt{2}A_{27}+\sqrt{2}A_8-\sqrt{5}C_8\)$          & $T + E$ \\
$D^0\to\ok\pi^0$     &$\frac{1}{10}\(3A_{27}-2A_{8}+\sqrt{10}C_{8}\)$                  & $(C - E)/\sqrt{2}$ \\
$D^0\to\ok\eta_8$    &$-\frac{1}{10\sqrt{3}}\left(3A_{27}-2A_{8}+\sqrt{10}C_{8}\right)$& $-(C - E)/\sqrt{6}$ \\
$D^+\to\ok\pi^+$     &$\frac{1}{\sqrt{2}}A_{27}$                                       & $T + C$ \\
$D^+_s\to\ok K^+$   &$\frac{1}{5}\(\sqrt{2}A_{27}+\sqrt{2}A_{8}+\sqrt{5}C_{8}\)$       & $C + A$ \\
$D^+_s\to\pi^+\eta_8$&$\frac{1}{5\sqrt{3}}\(-3A_{27}+2A_{8}+\sqrt{10}C_{8}\)$          & $-2(T - A)/\sqrt{6}$ \\ \hline
\multicolumn{3}{|c|}{$SU(3)_F$ octet-singlet final states} \\ \hline
$D^0\to \ok\eta_1$   &$-\frac{1}{\sqrt{15}}\(\sqrt{2}A_8-\sqrt{5}C_8\)$         & $-(C + 2E)/\sqrt{3}$ \\
$D^+_s\to\pi^+\eta_1$&$\frac{1}{\sqrt{15}}\(\sqrt{2}A_{8}+\sqrt{5}C_{8}\)$   & $(T + 2A)/\sqrt{3}$ \\ \hline
\end{tabular}
\end{center}
\caption{Amplitudes for CF $D\to PP$ processes using the $SU(3)_F$ matrix-elements and flavor-flow representations. We have separated decays to final states
involving octets only and those also involving singlets. Overall factors containing $G_F$ and $V_{\rm CKM}$, that are identical in both representations, have
been left out for brevity. \label{tab:CFeta8eta1}}
\end{table}

Focusing our attention, first, on the octet-octet final-state amplitudes in Table \ref{tab:CFeta8eta1}, we see that there are six decay amplitudes that depend on three
$SU(3)_F$ reduced matrix elements. Therefore, there must be three relationships between these amplitudes. They are,
\bea
\cA(D^0\to K^-\pi^+) + \sqrt{2}\cA(D^0\to\ok\pi^0) &=& \cA(D^+\to\ok\pi^+) ,~ \\
\cA(D^0\to\ok\pi^0) + \sqrt{3}\cA(D^0\to\ok\eta_8) &=& 0 ,~ \\
\sqrt{2}\,\cA(D^+\to\ok\pi^+) + \sqrt{3}\,\cA(D^+_s\to\pi^+\eta_8)  &=& \sqrt{2}\,\cA(D^+_s\to\ok K^+) .~
\eea
These relationships match with sum rules previously demonstrated in Ref.~\cite{Grossman:2012ry}.
Of these relationships, the first is a consequence of isospin symmetry, while the other two originate from the full $SU(3)_F$ symmetry. Note that these relationships are
satisfied by both matrix elements and diagrams. Although there are still four diagrams in play, every amplitude can be written in terms of three distinct linear combinations of them.
One can establish a one-to-one correspondence between the combinations of these diagrams and matrix elements as follows. The $SU(3)_F$ reduced matrix elements
can be expressed in terms of the flavor-flow diagrams using
\beq\label{eq:octets}
\[\bma A_{27} \\ A_{8} \\ C_{8} \ema\] ~=~
\[\bma 0 & \sqrt{2} & 0 \\
\frac{5\sqrt{2}}{4} & -\sqrt{2}&\frac{5\sqrt{2}}{4} \\
-\frac{\sqrt{5}}{2} & 0 &\frac{\sqrt{5}}{2} \ema\]\,
\[\bma T + E \\ T + C \\ C + A \ema\] .~
\end{equation}
Since the transformation matrix has a non-zero determinant it is invertible thus establishing a one-to-one correspondence. Next, we turn our attention to the octet-singlet final state
amplitudes in Table \ref{tab:CFeta8eta1}. Here, there are two decay amplitudes that depend on two $SU(3)_F$ reduced matrix elements and two combinations of flavor-flow diagrams.
Once again, the  reduced matrix elements can be expressed in terms of diagrammatic amplitudes using
\beq\label{eq:singlets}
\[\bma A_{8} \\ C_{8} \ema\] ~=~
\[\bma \frac{\sqrt{10}}{4} & \frac{\sqrt{10}}{4} \\
       -\frac{1}{2} & \frac{1}{2} \ema\]\,
\[\bma C + 2 E \\ T + 2 A \ema\] .~
\end{equation}
Here too, we see that the transformation matrix is invertible and a one-to-one correspondence exists.

Although Eqs.~(\ref{eq:octets}) and (\ref{eq:singlets}) establish a one-to-one correspondence between matrix elements and sets of flavor-flow amplitudes, it is easy to see that the
correspondences are not the same. On the $SU(3)_F$ matrix-elements side this can be traced back to the definitions: while the $\bm{27}$ appears only in
the $\bm{8}\times\bm{8}$ final states, the $\bm{8}$ appears in both $\bm{8}\times\bm{8}$ and $\bm{8}\times\bm{1}$. In principle, these final state octets are different and the
corresponding amplitudes should be treated as such. On the side of topological flavor-flow amplitudes, similarly, this implies distinct diagrams for octet-octet and octet-singlet
final states. In order to make these distinctions clear for the matrix elements, we use the following (re)definitions.
\bea
A_{27}    ~=~ \Braket{\bm{27}|\cO^{\bm{\overline{15}}}|\bm{3}}\,, \quad
A_8       ~=~ \Braket{\bm{8}_{\bm{8}\times\bm{8}}|\cO^{\bm{\overline{15}}}|\bm{3}}\,, \quad
C_8       ~=~ \Braket{\bm{8}_{\bm{8}\times\bm{8}}|\cO^{\bm{6}}|\bm{3}}\,, \label{eq:revSU3me8} \\
A^{(1)}_8 ~=~ \Braket{\bm{8}_{\bm{8}\times\bm{1}}|\cO^{\bm{\overline{15}}}|\bm{3}}\,, \quad
C^{(1)}_8 ~=~ \Braket{\bm{8}_{\bm{8}\times\bm{1}}|\cO^{\bm{6}}|\bm{3}}\,.~~~~~~~~~~~~~~ \label{eq:revSU3me1}
\eea
For diagrams, we simply add the subscript $1$ to represent the octet-singlet final states. Since these changes affect only the octet-singlet final states part of Table \ref{tab:CFeta8eta1},
we have listed the changes in Table \ref{tab:revCFeta8eta1}.
\begin{table}[h!]
\begin{center}
\begin{tabular}{|c|c|c|}\hline
Decay & Matrix Elements & Diagrams \\ \hline
\multicolumn{3}{|c|}{$SU(3)_F$ octet-singlet final states} \\ \hline
$D^0\to \ok\eta_1$   &$-\frac{1}{\sqrt{15}}\(\sqrt{2}A^{(1)}_8-\sqrt{5}C^{(1)}_8\)$     & $-(C_1 + 2E_1)/\sqrt{3}$ \\
$D^+_s\to\pi^+\eta_1$&$\frac{1}{\sqrt{15}}\(\sqrt{2}A^{(1)}_{8}+\sqrt{5}C^{(1)}_{8}\)$   & $(T_1 + 2A_1)/\sqrt{3}$ \\ \hline
\end{tabular}
\end{center}
\caption{Amplitudes for CF $D\to PP$ with octet-singlet final states using $SU(3)_F$ matrix-elements and diagrams. Overall factors containing $G_F$ and $V_{\rm CKM}$, that are identical in both representations, have been left out for brevity. \label{tab:revCFeta8eta1}}
\end{table}

Let us, now, reconsider the $\chi^2$ minimization fits presented in Section \ref{sec:fits} in light of the newly-defined amplitudes. The $SU(3)_F$ matrix-elements approach for the fits involved three complex-valued amplitudes ($A_{27}, A_8$, and $C_8$), rather than the five defined here ($A_{27}, A_8, C_8, A_8^{(1)}$, and $C_8^{(1)}$). The implicit assumptions in the fit were,
\beq
A_8^{(1)} ~=~ A_8~, \quad {\rm and} \quad C_8^{(1)} ~=~ C_8~.~
\eeq
The results of the fit were poor showing that matrix elements for the octet-octet and octet-singlet final states may not be identical. On the other hand, the diagrammatic approach involved four complex-valued amplitudes ($T, C, E$, and $A$), as opposed to eight ($T, C, E, A, T_1, C_1, E_1$, and $A_1$). The diagrammatic fits, therefore, assumed $X ~=~ X_1$ where $X = T, C, E$, and $A$.

In either scenario, matrix elements or diagrams, the parametrizations established in this section are insufficient by themselves to produce a reasonable fit.
Both parametrizations are equivalent as established above, and as such there are five complex-valued amplitudes which correspond to nine real-valued
parameters (five magnitudes and four relative phases). With only eight branching ratios to fit, lack of additional input will lead to overfitting. Clearly, additional
input is necessary to perform a fit.

On the $SU(3)_F$ matrix-elements side a fit was made possible by the assumption that the reduced matrix elements for octet-octet and octet-singlet final states
were the same. These assumptions put two complex constraints reducing the number of fit parameters to five. The resulting fit was rather poor. On the other hand,
the flavor-flow side assumption that individual diagrams corresponding to octet-octet and octet-singlet final states are the same led to four complex constraints
reducing the number of fit parameters to seven. The resulting fit was good.

Due to the established equivalence between the $SU(3)_F$ matrix-elements and topological flavor-flow approaches, one naturally inquires about the consequence
of either set of assumptions on the alternate parametrization. The $SU(3)_F$ matrix-elements side assumptions, $A_8^{(1)}=A_8$ and $C_8^{(1)}=C_8$, lead to the
following relationships on the topological flavor-flow side,
\bea
\[(T + C) + 5(E + A)\] - \sqrt{5}\[(T_1 + C_1) + 2(E_1 + A_1)\] &=& 0 ~,~~ \label{eq:toprel1} \\
\sqrt{5}\[(T - C) + (E - A)\] + \[(T_1 - C_1) - 2(E_1 - A_1)\]  &=& 0 ~.~~  \label{eq:toprel2}
\eea
Similarly, the flavor-flow side assumptions that $X ~=~ X_1$ where $X = T, C, E$, and $A$, lead to the number of reduced matrix elements being greater than that of the flavor-flow amplitudes. Then, the relations of Eqs.~(\ref{eq:octets}) and (\ref{eq:singlets}) lead to the following phenomenological relationship on the $SU(3)_F$ matrix-elements side,
\beq
 3\,A_{27} + 8\,A_8 - 4\sqrt{5}\,A_8^{(1)} ~=~ 0 ~.~ \label{eq:SU3rel}
\eeq
A fit performed with $A_{27}, C_8, A_8^{(1)}$, and $C_8^{(1)}$ as the available matrix elements, but with $A_8$ constrained
through the relationship in Eq.~(\ref{eq:SU3rel}), yields identical results as the diagrammatic fit assuming an equivalence between octet-octet and octet-singlet final states. We find,

\bea
\chi^2_{\rm min}/{\rm dof} &=& 1.36/1 ~, \nl
A_{27}    &=& (0.256 \pm 0.003)\,{\rm GeV}^3 ~,\nl
C_8       &=& (0.357 \pm 0.010)\,e^{i (44 \pm 1)^\circ}~{\rm GeV}^3~, \nl
A_8^{(1)} &=& (0.71  \pm 0.02 )\,e^{i (-67 \pm 1)^\circ}~{\rm GeV}^3~, \nl
C_8^{(1)} &=& (0.348 \pm 0.005)\,e^{i(-143 \pm 2)^\circ}~{\rm GeV}^3~.
\eea

Note that neither Eqs.~(\ref{eq:toprel1}) and (\ref{eq:toprel2}), nor Eq.~(\ref{eq:SU3rel}) automatically imply underlying relationships between the related parameters at a fundamental level. However, the fact that the fit on the diagrammatic side is far better than on the matrix-elements side, indicates a phenomenological preference for Eq.~(\ref{eq:SU3rel}).

\begin{table}[!htbp]
\begin{centering}
\begin{tabular}{|c|c|} \hline
Input relationship                        & $\chi^2_{\rm min}$ \\ \hline
$A_{27} ~=~ 0$                            & 2660    \\
$\arg(A_8^{(1)}) ~=~ \arg(A_8)$ and
$\arg(C_8^{(1)}) ~=~ \arg(C_8)$           &   175  \\
$\arg(A_8^{(1)}) ~=~ \arg(C_8^{(1)})$ and
$\arg(A_8) ~=~ \arg(C_8)$                 &  162    \\
$A_8 ~=~ 0$                               &  160    \\
$C_8 ~=~ 0$                               &   88.2  \\
$A_8^{(1)} ~=~ 0$                         &    9.31 \\
$A_8^{(1)} ~=~ A_8$                       &    8.32 \\
$C_8^{(1)} ~=~ 0$                         &    7.80 \\
$|A_8^{(1)}| ~=~ |A_8|$ and
$|C_8^{(1)}| ~=~ |C_8|$                   &    3.05  \\
$3\,A_{27} + 8\,A_8 - 4\sqrt{5}\,
A_8^{(1)} ~=~ 0$                          &    1.36 \\
$|A_8^{(1)}| ~=~ |C_8^{(1)}|$ and
$|A_8| ~=~ |C_8|$                         &    0.937 \\
$C_8^{(1)} ~=~ C_8$                       &    0.801  \\ \hline
\end{tabular}
\caption{Input relationships between $SU(3)_F$ matrix elements used to perform $\chi^2$ minimization fits, listed in descending
order of minimum $\chi^2$ value obtained in a fit. Each input relationship adds two real-valued constraints. The corresponding fits each
have one degree of freedom. $\arg(X)$ refers to the phase of the matrix element $X$.}\label{tab:input}
\end{centering}
\end{table}

For the sake of completeness, we performed additional seven-parameter $\chi^2$-minimization fits to the data,
each time changing the input relationship between the $SU(3)_F$ matrix elements. The minimum values of $\chi^2$
obtained in these fits are listed in Table~\ref{tab:input}. We see that all but one of the fits appear worse than the fit
with octet-octet and octet-singlet diagrams set equal to each other. The fit that has a smaller minimum $\chi^2$ is
one where we imposed the relationship $C_8^{(1)} = C_8$. For this fit, we find,

\bea
\chi^2_{\rm min}/{\rm dof} &=& 0.801/1 ,~
\nonumber \\
A_{27}   &=& (0.253 \pm 0.003) {\rm GeV}^3 ,~
\nonumber \\
A_8      &=& (1.021  \pm 0.009 )\,e^{i(96 \pm 1)^\circ}~{\rm GeV}^3 ,~
\nonumber \\
C_8      &=& (0.089 \pm 0.005)\,e^{i ( -55 \pm 5)^\circ}~{\rm GeV}^3 ,~
\nonumber \\
A_8^{(1)}&=& (0.79 \pm 0.02)\,e^{i (-140 \pm 1)^\circ}~{\rm GeV}^3 .~
\eea
Diagrammatically, this input relationship is equivalent to Eq.~(\ref{eq:toprel1}) on the flavor-flow side.

We conclude this section with the following observation. Since in the most general case the number of
basis decay parameters exceeds the number of experimentally-measured CF decay modes, additional
assumptions must be employed to extract individual reduced matrix elements or flavor-flow amplitudes
presented above. Yet, the relations such as Eqs.~(\ref{eq:octets}) and (\ref{eq:singlets}) are rather
general. This allows us to make a comment regarding hadronic final state interactions (FSI) in charm.
In the $SU(3)_F$ limit FSI cannot change the values of the reduced matrix elements. In other words,
action of the strong interaction $S$-matrix on the basis of the $SU(3)_F$ reduced matrix elements
leaves this basis invariant. This is not necessarily so for the individual flavor-flow amplitudes. Yet,
the {\it combinations} of these amplitudes are preserved under strong FSI. Extraction of the magnitudes
and phases of the individual amplitudes is only possible with additional assumptions.

\section{Conclusions}
\label{sec:conclusions}

Nonleptonic decays of charmed mesons provide plethora of interesting information about QCD
dynamics in its nonperturbative regime. In this paper we discussed two phenomenological
parametrizations of those decay amplitudes based on $SU(3)_F$ symmetry, which have been proven
equivalent in the decays of B-mesons.

We argue that application of such parametrizations to charm decays require care due to insufficient number of
experimentally-measured decay models and the presence of
final state interactions. Noting that the Wolfenstein parameter $\lambda$ is external to any QCD-based
parametrization of decay amplitudes, the equivalency of the flavor $SU(3)_F$ and the topological flavor-flow
approaches must be separately realized for the Cabibbo-favored decays of charmed mesons. Including
decays to the physical $\eta$ and $\eta^\prime$ mesons in our description, we find relationships between
the basis parameters of the flavor-flow amplitudes. This can be interpreted from the point of view that quark
rescatterings imply that only certain linear combinations of flow diagrams can contribute to the decay amplitudes.
We presented extractions of the basis amplitudes in two approaches under various assumptions.

\bigskip
{\bf Acknowledgments}: This work was financially supported in part by NSF Grant No. PHY2013984 (BB), PHY1915142
(AD \& JW), and the U.S. Department of Energy under contract de-sc0007983 (AAP). BB thanks J.~L. Rosner for useful conversations.

\bibliographystyle{JHEP}
\bibliography{ref_SU(3)amps}

\providecommand{\href}[2]{#2}\begingroup\raggedright\begin{thebibliography}{10}

\bibitem{Petrov:2021idw}
A.~A. Petrov, \emph{{Indirect Searches for New Physics}}. CRC Press, Boca
  Raton, 5, 2021,
  \href{https://doi.org/10.1201/9781351176019}{10.1201/9781351176019}.

\bibitem{Savage:1989ub}
M.~Savage and M.~Wise, \emph{{SU(3) predictions for nonleptonic B meson
  decays}}, \href{https://doi.org/10.1103/PhysRevD.39.3346}{\emph{Phys. Rev. D}
  {\bfseries 39} (1989) 3346}.

\bibitem{Grinstein:1996us}
B.~Grinstein and R.~Lebed, \emph{{SU(3) decomposition of two-body B decay
  amplitudes}}, \href{https://doi.org/10.1103/PhysRevD.53.6344}{\emph{Phys.
  Rev. D} {\bfseries 53} (1996) 6344}
  [\href{https://arxiv.org/abs/hep-ph/9602218}{{\ttfamily hep-ph/9602218}}].

\bibitem{He:2018joe}
X.-G. He, Y.-J. Shi and W.~Wang, \emph{{Unification of Flavor SU(3) Analyses of
  Heavy Hadron Weak Decays}},
  \href{https://doi.org/10.1140/epjc/s10052-020-7862-5}{\emph{Eur. Phys. J. C}
  {\bfseries 80} (2020) 359}
  [\href{https://arxiv.org/abs/1811.03480}{{\ttfamily 1811.03480}}].

\bibitem{He:2018php}
X.-G. He and W.~Wang, \emph{{Flavor SU(3) Topological Diagram and Irreducible
  Representation Amplitudes for Heavy Meson Charmless Hadronic Decays: Mismatch
  and Equivalence}},
  \href{https://doi.org/10.1088/1674-1137/42/10/103108}{\emph{Chin. Phys. C}
  {\bfseries 42} (2018) 103108}
  [\href{https://arxiv.org/abs/1803.04227}{{\ttfamily 1803.04227}}].

\bibitem{Quigg:1979ic}
C.~Quigg, \emph{{Charmed Meson Decays and the Structure of the Charged Weak
  Current}}, \href{https://doi.org/10.1007/BF01477308}{\emph{Z. Phys. C}
  {\bfseries 4} (1980) 55}.

\bibitem{Kaeding:1995vq}
T.~A. Kaeding, \emph{{Tables of SU(3) isoscalar factors}},
  \href{https://doi.org/10.1006/adnd.1995.1011}{\emph{Atom. Data Nucl. Data
  Tabl.} {\bfseries 61} (1995) 233}
  [\href{https://arxiv.org/abs/nucl-th/9502037}{{\ttfamily nucl-th/9502037}}].

\bibitem{Savage:1991wu}
M.~Savage, \emph{{SU(3) violations in the nonleptonic decay of charmed
  hadrons}}, \href{https://doi.org/10.1016/0370-2693(91)91917-K}{\emph{Phys.
  Lett. B} {\bfseries 257} (1991) 414}.

\bibitem{Gronau:1994rj}
M.~Gronau, O.~Hernandez, D.~London and J.~Rosner, \emph{{Decays of B mesons to
  two light pseudoscalars}},
  \href{https://doi.org/10.1103/PhysRevD.50.4529}{\emph{Phys. Rev. D}
  {\bfseries 50} (1994) 4529}
  [\href{https://arxiv.org/abs/hep-ph/9404283}{{\ttfamily hep-ph/9404283}}].

\bibitem{Buras:1998ra}
A.~J. Buras and L.~Silvestrini, \emph{{Nonleptonic two-body B decays beyond
  factorization}},
  \href{https://doi.org/10.1016/S0550-3213(99)00712-9}{\emph{Nucl. Phys. B}
  {\bfseries 569} (2000) 3}
  [\href{https://arxiv.org/abs/hep-ph/9812392}{{\ttfamily hep-ph/9812392}}].

\bibitem{Imbeault:2006ss}
M.~Imbeault, A.~Datta and D.~London, \emph{{Hadronic B decays: A General
  approach}}, \href{https://doi.org/10.1142/S0217751X07036397}{\emph{Int. J.
  Mod. Phys. A} {\bfseries 22} (2007) 2057}
  [\href{https://arxiv.org/abs/hep-ph/0603214}{{\ttfamily hep-ph/0603214}}].

\bibitem{Bhattacharya:2012pc}
B.~Bhattacharya, M.~Gronau and J.~L. Rosner, \emph{{Nonleptonic Charm Decays
  and CP Violation}},  in \emph{{5th International Workshop on Charm Physics}},
  7, 2012, \href{https://arxiv.org/abs/1207.6390}{{\ttfamily 1207.6390}}.

\bibitem{Bhattacharya:2012ah}
B.~Bhattacharya, M.~Gronau and J.~L. Rosner, \emph{{CP asymmetries in
  singly-Cabibbo-suppressed $D$ decays to two pseudoscalar mesons}},
  \href{https://doi.org/10.1103/PhysRevD.85.054014}{\emph{Phys. Rev. D}
  {\bfseries 85} (2012) 054014}
  [\href{https://arxiv.org/abs/1201.2351}{{\ttfamily 1201.2351}}].

\bibitem{Zeppenfeld:1980ex}
D.~Zeppenfeld, \emph{{SU(3) relations for B meson decays}},
  \href{https://doi.org/10.1007/BF01429835}{\emph{Z. Phys. C} {\bfseries 8}
  (1981) 77}.

\bibitem{Muller:2015lua}
S.~M\"uller, U.~Nierste and S.~Schacht, \emph{{Topological amplitudes in $D$
  decays to two pseudoscalars: A global analysis with linear $SU(3)_F$
  breaking}}, \href{https://doi.org/10.1103/PhysRevD.92.014004}{\emph{Phys.
  Rev. D} {\bfseries 92} (2015) 014004}
  [\href{https://arxiv.org/abs/1503.06759}{{\ttfamily 1503.06759}}].

\bibitem{Wolfenstein:1983yz}
L.~Wolfenstein, \emph{{Parametrization of the Kobayashi-Maskawa matrix}},
  \href{https://doi.org/10.1103/PhysRevLett.51.1945}{\emph{Phys. Rev. Lett.}
  {\bfseries 51} (1983) 1945}.

\bibitem{Golden:1989qx}
M.~Golden and B.~Grinstein, \emph{{Enhanced CP Violations in Hadronic Charm
  Decays}}, \href{https://doi.org/10.1016/0370-2693(89)90353-5}{\emph{Phys.
  Lett. B} {\bfseries 222} (1989) 501}.

\bibitem{Pirtskhalava:2011va}
D.~Pirtskhalava and P.~Uttayarat, \emph{{CP Violation and Flavor SU(3) Breaking
  in D-meson Decays}},
  \href{https://doi.org/10.1016/j.physletb.2012.04.039}{\emph{Phys. Lett. B}
  {\bfseries 712} (2012) 81} [\href{https://arxiv.org/abs/1112.5451}{{\ttfamily
  1112.5451}}].

\bibitem{Cheng:2012wr}
H.-Y. Cheng and C.-W. Chiang, \emph{{Direct CP violation in two-body hadronic
  charmed meson decays}},
  \href{https://doi.org/10.1103/PhysRevD.85.034036}{\emph{Phys. Rev. D}
  {\bfseries 85} (2012) 034036}
  [\href{https://arxiv.org/abs/1201.0785}{{\ttfamily 1201.0785}}].

\bibitem{Cheng:2012xb}
H.-Y. Cheng and C.-W. Chiang, \emph{{SU(3) symmetry breaking and CP violation
  in D -\ensuremath{>} PP decays}},
  \href{https://doi.org/10.1103/PhysRevD.86.014014}{\emph{Phys. Rev. D}
  {\bfseries 86} (2012) 014014}
  [\href{https://arxiv.org/abs/1205.0580}{{\ttfamily 1205.0580}}].

\bibitem{Feldmann:2012js}
T.~Feldmann, S.~Nandi and A.~Soni, \emph{{Repercussions of Flavour Symmetry
  Breaking on CP Violation in D-Meson Decays}},
  \href{https://doi.org/10.1007/JHEP06(2012)007}{\emph{JHEP} {\bfseries 06}
  (2012) 007} [\href{https://arxiv.org/abs/1202.3795}{{\ttfamily 1202.3795}}].

\bibitem{Hiller:2012xm}
G.~Hiller, M.~Jung and S.~Schacht, \emph{{SU(3)-flavor anatomy of nonleptonic
  charm decays}}, \href{https://doi.org/10.1103/PhysRevD.87.014024}{\emph{Phys.
  Rev. D} {\bfseries 87} (2013) 014024}
  [\href{https://arxiv.org/abs/1211.3734}{{\ttfamily 1211.3734}}].

\bibitem{Falk:2001hx}
A.~F. Falk, Y.~Grossman, Z.~Ligeti and A.~A. Petrov, \emph{{$SU(3)$ breaking
  and $D^0 - \bar D^0$ mixing}},
  \href{https://doi.org/10.1103/PhysRevD.65.054034}{\emph{Phys. Rev. D}
  {\bfseries 65} (2002) 054034}
  [\href{https://arxiv.org/abs/hep-ph/0110317}{{\ttfamily hep-ph/0110317}}].

\bibitem{Datta:2001ir}
A.~Datta, H.~J. Lipkin and P.~J. O'Donnell, \emph{{Simple relations for
  two-body B decays to charmonium and tests for eta eta-prime mixing}},
  \href{https://doi.org/10.1016/S0370-2693(02)01247-9}{\emph{Phys. Lett. B}
  {\bfseries 529} (2002) 93}
  [\href{https://arxiv.org/abs/hep-ph/0111336}{{\ttfamily hep-ph/0111336}}].

\bibitem{Petrov:1997yf}
A.~A. Petrov, \emph{{Intrinsic charm of light mesons and CP violation in heavy
  quark decay}}, \href{https://doi.org/10.1103/PhysRevD.58.054004}{\emph{Phys.
  Rev. D} {\bfseries 58} (1998) 054004}
  [\href{https://arxiv.org/abs/hep-ph/9712497}{{\ttfamily hep-ph/9712497}}].

\bibitem{Zyla:2020zbs}
{\scshape Particle Data Group} collaboration, \emph{{Review of Particle
  Physics}}, \href{https://doi.org/10.1093/ptep/ptaa104}{\emph{PTEP} {\bfseries
  2020} (2020) 083C01}.

\bibitem{Bhattacharya_2010}
B.~Bhattacharya and J.~L. Rosner, \emph{Charmed meson decays to two
  pseudoscalars},
  \href{https://doi.org/10.1103/physrevd.81.014026}{\emph{Physical Review D}
  {\bfseries 81} (2010) }.

\bibitem{Grossman:2012ry}
Y.~Grossman and D.~J. Robinson, \emph{{SU(3) Sum Rules for Charm Decay}},
  \href{https://doi.org/10.1007/JHEP04(2013)067}{\emph{JHEP} {\bfseries 04}
  (2013) 067} [\href{https://arxiv.org/abs/1211.3361}{{\ttfamily 1211.3361}}].

\end{thebibliography}\endgroup

\end{document}